\documentclass[prl,aps,superscriptaddress,showpacs,amsmath,twocolumn]{revtex4}

\usepackage[dvips]{graphicx}
\usepackage{dcolumn}
\usepackage{bm}
\usepackage{amsmath}

\begin{document}




\title{ \emph{Ab-initio} molecular dynamics for high-pressure liquid Hydrogen
} 

\author{Sandro Sorella}
\affiliation{Democritos National Simulation Center, 
SISSA, Via Beirut n.2, Trieste, Italy} 
\author{ Claudio Attaccalite}
\affiliation{Institute for Electronics, Microelectronics, and Nanotechnology Dept.  ISEN B.P. 60069 59652 Villeneuve d'Ascq Cedex France} 
\date{\today}
\begin{abstract}
We introduce an efficient scheme for the molecular dynamics 
 of electronic systems by means of  quantum Monte Carlo. 
The evaluation of 
the (Born-Oppenheimer) forces acting on the ionic positions 
is achieved by two main ingredients: i) 
the forces are computed with finite and small variance,  which allows  
the simulation of a  a  large number of atoms, 
ii) the statistical noise 
corresponding to the forces is used to drive the dynamics 
at finite temperature by means of  an appropriate friction matrix. 
A first application to the high-density phase of Hydrogen is 
given, supporting the stability of the  
 liquid phase  at $\simeq 300GPa$ and $\simeq 400K$. 
\end{abstract}
\pacs{47.11.Mn, 02.70.Ss, 61.20.Ja, 62.50.+p}

\maketitle

\narrowtext
The phase diagram of Hydrogen at high pressure is still under intense 
study from the experimental and theoretical point of view.
In particular in the low temperature high-pressure regime there is yet no clear evidence of a metallic atomic solid, and a suggestion was given in Ref.(\cite{galli}) that the liquid phase is instead more  stable.  
Indeed, for  high pressures around $300GPa$,  
a two fluid (proton and electron) superconducting  phase of the conventional 
type, namely induced by the  strong electron-phonon coupling, 
has been conjectured\cite{ashcroft}.
In this work we use an improved  {\em ab}-initio molecular dynamics (AMD) 
by using accurate forces computed by Quantum Monte Carlo (QMC).   
We present preliminary 
results, showing  that the   liquid phase is energetically more stable, due to the strong electron correlation, at least within the Resonating Valence Bond 
(RVB)  variational approach\cite{vanilla}, which is very accurate also in the solid phase.

AMD is well established as a powerful tool to investigate many-body condensed matter systems. 
Indeed, previous attempts to apply  Quantum Monte Carlo (QMC) 
 for   the dynamics of ions\cite{mitas} 
or for their thermodynamic properties\cite{penalty} are known, but they 
were limited to small number  $N$ of electrons  or to 
total energy corrections  
of the  AMD trajectories, namely without the explicit 
calculation of the forces. 

\vskip 2mm \noindent {\it Calculation of forces with finite variance.}\hskip 2mm
The simplest method  for accurate calculations within QMC, is given by 
the so called variational Monte Carlo (VMC), which allows to compute the 
variational energy expectation value $E_{VMC}={ \langle \psi_T |H | \psi_T \rangle \over \langle \psi_T | \psi_T \rangle }  $ 
of a highly accurate correlated wave function (WF)  $\psi_T$ by means of 
a statistical approach:
electronic configurations $\left\{ x \right\}$, with given 
electron positions $\vec r_i$  and spins $\sigma_i=\pm 1/2$ for $i=1,\cdots N$, 
 are usually generated 
by the Metropolis algorithm according to the probability density 
$\mu_x \propto \psi_T(x)^2$. Then $E_{VMC}$ is computed by averaging 
statistically over $\mu_x$ 
the so called local energy $e_L(x)= { \langle  \psi_T |H 
| x \rangle  \over \langle \psi_T | x \rangle } $, namely 
 $E_{VMC} = \int d\mu_x~ e_L(x)$, 
where $\int d \mu_x $ indicates conventionally the $3 N$ multidimensional 
integral over the electronic coordinates weighted by $\psi_T^2(x)$.
In the present work we assume that the WF  $\psi_T(x)= \langle x |
\psi_T \rangle=  J  \times 
 \det A$ is given by a correlated Jastrow factor $J$ 
times a determinant $D$ of a $N \times N$ matrix $A$, such 
as for instance a Slater determinant. 
The main ideas  of this approach can be straightforwardly generalized to more 
complicated  and more accurate WF's. 

The efficient calculation of the energy derivatives, namely  the forces 
 $ \vec f_{\vec R_i}  =- { \partial E_{VMC}  \over  \partial \vec R_i }$, 
for $i=1,\cdots N_A$, where $N_A$ is the number of atoms, 
is the most important ingredient for the AMD.
Within VMC they can be computed by simple differentiation 
of $E_{VMC}$, using that not only the Hamiltonian $H$ but also  
$\psi_T$ depend explicitly on the  
atomic positions $\vec R_i$. This leads to two different  contributions 
to the 
force $\vec f_{\vec R_i} = \vec f^{HF}_{\vec R_i} + \vec f^P_{\vec R_i}$,  
the Hellmann-Feynman  $\vec f^{HF}$ and the Pulay one 
$\vec f^P_{\vec R_i}$,  where:
\begin{eqnarray}
\vec f_{\vec R_i}^{HF} & = & -\int d\mu_x ~    \langle x | 
\partial_{\vec R_i} H | x \rangle  \label{HF}   \\
\vec f_{\vec R_i}^{P} & = & -2 \int d\mu_x ~ (e_L(x) -E_{VMC})   
\partial_{\vec R_i} {\rm log}|\psi_T (x)|   \label{P}  
\end{eqnarray}
However in order to obtain a  statistically meaningful average, namely with 
finite variance, 
some manipulations  are  necessary because the first integrand  
 may diverge when the atoms are close to some 
electronic positions, whereas the second integrand is analogously 
unbounded  when a  configuration $x$  
approaches the nodal surface   determined by $\psi_T(x)=0$. 
By defining   with  $d$ ($\delta$) 
 the distance of $x$ from the nodal region (the minimum 
electron-atom distance), 
 $e_L(x),  \partial_{\vec R_i} {\rm log}  \psi_T (x)  \simeq 1/d$ 
( $\langle x | \partial_{\vec R_i}  H | x \rangle  \simeq 1/\delta^2$), whereas 
$\mu_x \simeq d^2$ ($\mu_x \simeq \delta^2$), 
 leading to an unbounded integral of the 
 square integrand in Eq.(\ref{P}) (Eq.\ref{HF}),
 namely to {\em infinite variance}.  
The infinite variance problem in Eq.(\ref{HF}) was solved in several 
ways. Here we adopt a very elegant and efficient scheme 
proposed by Caffarel and Assaraf\cite{caffarel}. 
Instead the infinite variance problem in 
Eq.(\ref{P}) was not considered so  far, 
 and this is clearly a problem for  a meaningful 
definition of ionic   AMD consistent with QMC forces.

In this letter we solve this problem in the 
following simple way, by  using the so called re-weighting method.
We use a different probability distribution 
$\mu^\epsilon_x \propto \psi_G(x)^2$, determined by a guiding 
function $\psi_G(x)$:
\begin{equation}
\psi_G(x) = R^\epsilon (x) ( \psi_T(x)/R(x)) 
\end{equation}
where $R (x) \propto \psi_T(x) \to 0$  for $d \to 0$ 
is a ''measure'' of the distance from  the nodal surface $\psi_T(x)=0$.
By assumption $\psi_T$ may vanish  only when  $\det A=0$  
($J>0$) and therefore $R(x)$ is  chosen to depend only on $A$.  
 For reasons that will become clear later on we have adopted  the following 
expression:
\begin{equation}
R(x) = 1/\sqrt{ \sum\limits_{i,j=1}^N  |A^{-1}_{i,j}|^2}. \label{defr}
\end{equation}
Then the guiding function is defined  by properly regularizing $R(x)$, 
namely:
\begin{equation} \label{reg}
R^\epsilon(x) = \left\{ 
\begin{array}{cc}
R(x) & {\rm if ~}  R(x) \ge \epsilon \\
\epsilon (R(x)/\epsilon)^{ R(x)/\epsilon}  & {\rm if ~}  R(x) < \epsilon \\
\end{array} 
\right..
\end{equation}
The non obvious regularization for $ R(x) < \epsilon$ 
instead of e.g. 
$R^\epsilon (x)=Max[\epsilon, R(x)]$ was considered in order to 
satisfy the continuity of the first derivative of $\psi_G(x)$ when 
$R(x)=\epsilon$, thus ensuring that $\psi_G(x)$ remains as close as 
possible to the trial function $\psi_T$.   
In this way the Metropolis algorithm can be applied for generating 
configurations  according to a slightly different 
 probability  $\mu^\epsilon(x)$ 
and the exact expression 
of $\vec f^{P}_{\vec R_i}$ can be obtained by the so called umbrella 
average:
\begin{equation} \label{umbrella}
\vec f_{\vec R_i}^{P}  = {  -2 \int d\mu^\epsilon_x ~ S(x) (e_L(x) -E_{VMC})   
\partial_{\vec R_i} log \psi_T (x)   
\over \int d\mu^\epsilon_x ~ S(x) }.
\end{equation}
Now,  
 the re-weighting factor $S(x)= (\psi_T(x)/\psi_G(x))^2=
 Min \left[ 1, (R(x)/\epsilon)\right]^{2 -2 R(x)/\epsilon}  \propto d^2 $, 
cancels out the divergence of the integrand, that was instead present in 
Eq.(\ref{P}).
Hence the mentioned integrands  in the numerator 
and $S(x)$ ( $  \le 1$)  in the denominator 
of Eq.(\ref{umbrella}) represent  bounded random variables and 
have obviously {\em finite variance}.
In this way  the problem of infinite variance is definitely 
solved within this simple re-weighting scheme. 
Moreover, in the present method  $R(x)$ 
is not related to  an overall  factor of the  total WF, 
such as the total determinant, defined in  a very wide  range 
of values  $\simeq e^{ \propto  N } $ 
 over the various configurations. It is instead 
obtained by using  a  
quantity $R(x) \simeq {1 \over N}$ 
with small fluctuations. 
Therefore the present scheme is particularly 
efficient and stable also for large $N$. 

\begin{figure}[hbt]
\includegraphics[width=0.5\textwidth,angle=0]{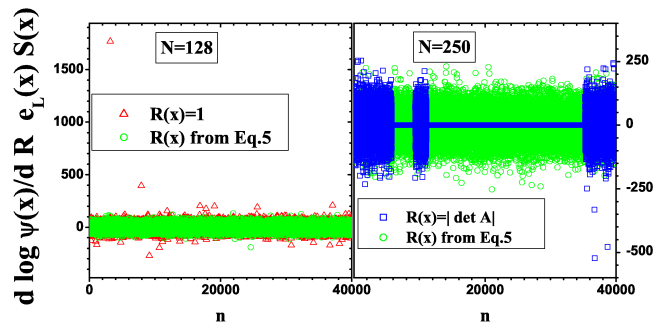}
\begin{center}
\caption{(color online). 
Evolution of the integrand in Eq.(\ref{umbrella}) as a function 
of the Monte Carlo iterations. Each new sample  is obtained after  
$2 N$ Metropolis trials.}
\label{fig1}
\end{center}
\end{figure}

We show in Fig.(\ref{fig1}) a comparison of several  methods for  
computing the Pulay force component acting on a   
Hydrogen proton at $r_s=1.31$ in a bcc lattice. 
As it is clear in the plot for the $N=128$ case, 
the difference between a method with finite 
variance 
and the standard one with infinite variance is evident. 
Moreover for $N=250$  the  
  simpler choice  $R(x)=|det A|$ 
 with  finite (but large) 
variance is clearly  very inefficient 
due to the difficulty to cross from the region with $R(x)<\epsilon$,
where the integrand almost vanishes (see Fig.\ref{fig1}), 
to the one with $R(x)> \epsilon$  
and viceversa. 
Instead  in the present scheme  an appropriate choice of $\epsilon$, such that 
 $ \int d\mu_x  S(x) \simeq 1/2$,   allows frequent barrier 
crossings any few Metropolis steps also for large $N$.
In this way 
 the $3 N_A \times 3 N_A$ correlation  matrix 
$\bar \alpha_{QMC}$,  defining  the statistical correlation 
between the force components, can be efficiently evaluated: 
\begin{equation}\label{noiseqmc}
\bar  \alpha_{QMC} (\vec R) =  <   (\vec f_{\vec R_i} - < \vec f_{\vec R_i} >)  
( \vec  f_{\vec R_j} -< \vec f_{\vec R_j}> )  >  
\end{equation}
where   the brackets $< >$ 
 indicate the statistical average over the QMC samples.
The correlation matrix  $\bar \alpha_{QMC}$, that  
within the conventional  method is not  even defined, 
will be a fundamental ingredient 
for a consistent AMD  with QMC forces and therefore the 
solution of the infinite variance problem 
 is particularly  important for this purpose.

 \vskip 2mm \noindent {\it  Langevin dynamics.} \hskip 2mm
In the following derivation we  assume that ions have 
unit mass, that can be generally  obtained by e.g. 
a simple rescaling of lengths for each ion independently. 
 For clarity and compactness of notations,
we also  omit the ionic subindices $i$  when  
not explicitly necessary.
Moreover   matrices (vectors) 
are indicated by a bar (arrow)  over the corresponding 
symbols, and  the matrix-vector product is also implicitly understood. 
We start therefore by the   following 
  AMD  equations for the 
ion coordinates $\vec R$ and velocities $\vec v$:
\begin{eqnarray} \label{ldyn}
 \dot {\vec v }  &=&  -  \bar \gamma(\vec R) 
  \vec v   +   \vec f(\vec R)   +\vec \eta  (t)   \\
 \dot{  \vec R }   &=&  \vec v    \label{dynR} 
\end{eqnarray} 
By using the fluctuation-dissipation theorem 
the friction matrix $\bar \gamma$ is related to the temperature 
$T$ (henceforth the Boltzmann constant $k_B=1$) by the relation:
\begin{equation} \label{eqpar}
\bar  \gamma(\vec R) = { 1 \over 2 T } \bar  \alpha (\vec R)
\end{equation}
where  $ \bar \alpha (\vec R)$ is generally a 
 symmetric correlation matrix:
\begin{equation}
<\vec \eta_i(t) \vec \eta_j(t^\prime)>=\delta(t-t^\prime) \bar \alpha (\vec R). 
\end{equation}
It is important to emphasize that, as a remarkable generalization 
of the standard AMD  used in \cite{parrforce}, 
in the present approach  the 
friction matrix $\bar \gamma$, 
may depend explicitly on the ion positions $\vec R$, 
so that Eq.(\ref{eqpar}) can be satisfied 
even for a generic correlation matrix $\bar \alpha (\vec R)$. 
In fact, since forces are computed by QMC, we can assume 
that there exists  
also a QMC contribution to $\bar \alpha (\vec R)$:
\begin{equation} \label{forces}
\bar \alpha (\vec R)  = \bar \alpha_{0} + \Delta_0 \, 
 \bar \alpha_{QMC} (\vec R)  
\end{equation}
where $\Delta_0>0$ and 
$\bar \alpha_{0}$ is the identity 
matrix $\bar I$  up to another  positive constant $\alpha_0$,
$\bar \alpha_{0} = \alpha_0 \bar I $. 

\vskip 2mm \noindent {\it Integration of the Langevin dynamics.} \hskip 2mm
 In the interval $ t_n-{ \Delta t \over 2} 
  < t < t_n+{ \Delta t \over 2}$, for $\Delta t$ small, 
the positions 
$\vec R$ are changing a little and, within a good approximation, 
 we can neglect the $\vec R$ 
dependence in the RHS of Eq.(\ref{ldyn}). 
Moreover the velocities 
$v_n$ are computed at half-integer  times $t_n-{ \Delta t \over 2}$, whereas 
coordinates $\vec R_n$ are assumed to be defined at integer times 
$\vec R_n= \vec R(t_n)$. 
Then the integration of Eq.(\ref{ldyn}) in the mentioned intervals 
 can be recasted  in the following useful  form, 
where the force components are corrected by appropriate noisy vectors:
\begin{eqnarray} \label{MD}
\vec v_{n+1}  &=&e^{ - \bar \gamma \Delta t }  \vec v_{n} +
\bar \Gamma  ( \vec f(\vec R_n) +   \vec { \tilde \eta} )   \\
\vec R_{n+1} & =& \vec R_n+  \Delta t \,\vec v_{n+1} +O(\Delta^3) \label{eqappr}\\
\bar \Gamma &=& \bar \gamma^{-1} ( 1 - e^{ -\bar \gamma \Delta t } )  \\ 
{\vec {\tilde \eta} } &=& 
{ \bar \gamma  \over  2  \sinh( {  \Delta t \over 2}  \bar \gamma) } 
\int\limits_{t_n- {\Delta t \over 2} }^{t_n+{ \Delta t \over 2} } dt e^{ \bar \gamma (t-t_n)} \vec \eta (t)    \label{totnoise} 
\end{eqnarray}
By using  that
$\bar \alpha = 2 T  \bar \gamma$ from Eq.(\ref{eqpar}),  
 the correlator defining the discrete (time integrated) 
noise  $ \tilde {\vec \eta} $ can be computed explicitly and is given by:
\begin{equation} \label{totalnoise}
<  \vec {\tilde \eta}_i \vec {\tilde \eta}_j > = 
{ 2 T } \bar \gamma^2  {  \sinh ( \Delta t \bar \gamma  
) \over
 4 \sinh(  { \Delta t  \over 2} \bar \gamma)^2 }= \bar \alpha^{\prime} 
\end{equation}
This means that 
 the QMC noise has to be corrected in a non trivial way as explained in the 
following.   

\begin{figure}[hbt]
\includegraphics[width=8.cm,angle=0]{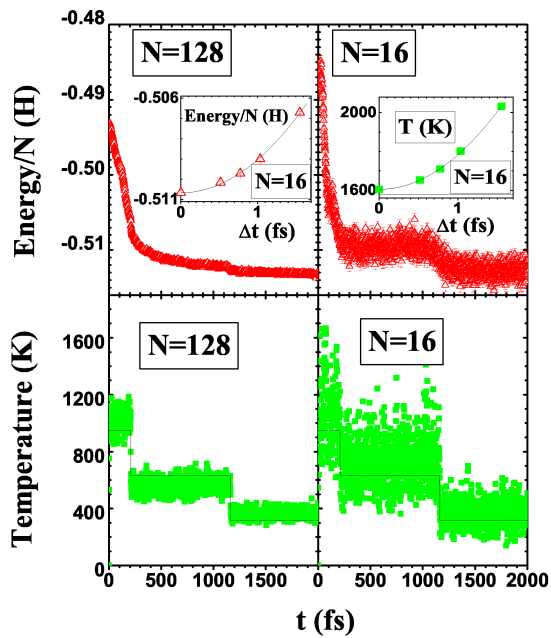}
\begin{center}
\caption{(color online). Evolution of the internal energy and temperature 
vs the AMD with QMC forces.
$\Delta t= \Delta_0=1.036fs$, 
$\alpha_0=0.7 k_B T a.u.$. Bottom: points represent instantaneous temperatures 
estimated by the average kinetic energy, lines represent the target 
 temperatures. They should coincide on average for $\Delta t \to 0$. 
See right inset for a target temperature of $1580K$. 
 The left inset shows  the corresponding energy.   
 The average energy, pressure and temperature in the last 0.5ps are
$-0.51319 \pm 0.00003 H$  ($-0.5127\pm 0.0001 H$)  
$364K \pm 5K$  ($364 \pm 10K$)   and 
$335\pm 2 GPa$ ($394\pm5 GPa$)  for $N=128$ ($N=16$), 
respectively. At each iteration the statistical noise on the total 
energy is $\simeq 5000K$.}
\label{fig2}
\end{center}
\end{figure}

\vskip 2mm  \noindent {\it Noise correction.} \hskip 2mm
The QMC noise is 
given during the simulation, and therefore in order to follow the 
correct dynamics another noise $\vec \eta^{\,ext}$ has to be added
to the noisy force components in a way  that the total integrated noise 
is the correct expression (\ref{totalnoise}), i.e.  
$ { \vec {\tilde  \eta} } = { \vec \eta}^{\,ext}+ \vec \eta_{QMC}.$
By using that the QMC noise 
in  Eq.(\ref{noiseqmc}) is obviously 
independent of the external noise, we easily obtain  
 the corresponding correlation matrix:
\begin{equation} \label{extnoise}
< \vec  \eta^{\,ext}_i \vec \eta^{\,ext}_j > = 
  \bar \alpha^{\prime} -\bar \alpha_{QMC} 
\end{equation}
On the other hand, after substituting the expression
  (\ref{forces}) in Eq.(\ref{eqpar}) 
$\bar \gamma={ 1\over 2 T }
 (\bar \alpha_0 + \Delta_0 \,\bar \alpha_{QMC})$
and using   the expression  (\ref{totalnoise}) for  
$\bar \alpha^\prime$, we finally  obtain  a positive definite matrix 
in Eq.(\ref{extnoise}) for $\Delta t\le \Delta_0$.\cite{proof} 
Hence $\vec \eta^{\,ext}$ is  
a generic Gaussian correlated noise 
that can be easily  sampled  by   standard algorithms. After that  
 the   random vector $\vec \eta^{\,ext}$     is   added    
 to the force $\vec f + \vec  \eta_{QMC}$ 
obtained by QMC,
and  replaces $\vec f + \vec { \tilde  \eta}  $  in Eq.(\ref{MD}). 
 This  finally allows to obtain 
an accurate AMD  with a corresponding 
small time step error. 
In our approach the choice $\alpha_0=0$ is also allowed  but,  
following Refs.(\onlinecite{parrforce},\onlinecite{mauri}),
much better performances of the AMD are  obtained with non zero $\alpha_0>0$ 
 and/or $\Delta_0> \Delta t$, namely with an external noise larger than 
the smallest  possible one ($\alpha_0=0,\Delta_0=\Delta t$). 
The main advantage of this technique is that, at each iteration, 
by means of 
Eq.(\ref{eqpar}),
the statistical noise on the total energy 
 and forces (see Fig.\ref{fig2})
  can be much larger than the target temperature
 $T$, 
and this allows to improve the QMC efficiency 
by several orders of magnitude.
\vskip 2mm \noindent {\it Optimization of  the WF.}  \hskip 2mm
In the following examples we use a variational WF $J \times \det A$ 
  that 
is able to provide a very accurate description of the
correlation energy, due to 
a particularly efficient  
choice of the determinant factor, that allows to describe the RVB
 correlations.\cite{casulamol,benzenedim}
The WF contains  several variational parameters, indicated by a 
vector  $\vec \beta $, 
 that have to be consistently optimized during the AMD. 
The Jastrow factor $J$ used here  depends  both on the charge and spin 
densities, for a total of  $\simeq N_A^2$ 
variational parameters.
We employ periodic boundary conditions; 
for each proton we use two  periodic gaussians centered 
at each  ionic position, 
whereas for the Jastrow we use only one Gaussian. 
As it is shown in the table, the accuracy of our WF is 
remarkable.
Indeed the small difference between the so called DMC
-providing the lowest possible  variational energy within
the same nodal surface of $\psi_T$-
 and the VMC energies 
clearly supports the accuracy of our calculation, as well as that the 
$N=54$ is an accidental closed shell, and will not be considered in the 
furthcoming  analisys.
\begin{table} \caption{\label{htab} 
Comparison of the total energy per proton (Hartree) 
 for Hydrogen  in the bcc lattice at $r_s=1.31$ compared with the 
  published ones with lowest energy (to our knowledge). 
All energies are in Hartree.}  
\begin{center}
\begin{tabular}{|l|c|c|c|r|}
\hline
 N & $E_{VMC}/N_A$ & $E_{VMC}/N_A$\cite{pier} & 
 $E_{DMC}/N_A $ &  $E_{DMC}/N_A $\cite{pier}  \\
\hline 
 16  & -0.48875(5) &  -0.4878(1) &  -0.49164(4)  & -0.4905(1) \\
 54  & -0.53573(2) &  -0.5353(2) & -0.53805(4)  & -0.5390(5) \\
 128 & -0.49495(1) & -0.4947(2) &  -0.49661(3)  & -0.4978(4)  \\
 250 & -0.49740(2)  & -  & -0.49923(2) & - \\ 
\hline
\hline
\end{tabular}
\end{center}
\end{table}

In order to optimize the WF we use the recent 
method introduced in Ref.\onlinecite{hesscyrus}, devised here 
in an appropriate way to optimize a large number 
of parameters during the AMD simulation. 
At each iteration time $t_n$ we compute the  generalized energy 
gradients 
$\vec g_n = \bar s^{-1} {  \partial E_{VMC}  \over \partial \vec \beta } $ 
where $ \hat s$ is the reduced 
overlap matrix between the logarithmic WF derivatives,
appropriately regularized as in Ref.\onlinecite{benzenedim}.
Then we use the mentioned method to minimize  the energy 
in the linear space  spanned by $p-$ vectors 
$\vec g_{n},\vec g_{n-1},\cdots \vec g_{n-p+1}$ 
corresponding to the  previous $p$ iterations ($p\simeq 
100$ for large systems).
The ionic positions are then consistently updated according to Eq.(\ref{MD})
and, in order to improve the QMC stability, the corresponding change 
in the electronic variational parameters is reduced  by  a factor four.
This factor clearly leads to  a 
slowing  down of the electron dynamic, that however 
turns out to  remain  very 
close to the Born-Oppenheimer one,  even by   
performing only one step of  WF optimization (with $p\simeq 100$) 
each time  the  ionic positions are changed. 
On the other hand, as shown in the insets of Fig.(\ref{fig2}), 
 the small  bias due to the finite time discretization 
$\Delta t$ appears  to produce only an effective change  of the average 
temperature (calculated by the average ion kinetic energy) and corresponding 
consistent change  of the internal energy (see left inset).

\vskip 2mm \noindent {\it Application to high-pressure Hydrogen.} \hskip 2mm
The phase diagram of Hydrogen is still under debate especially for
a possible stable low temperature $\simeq 400K$
high pressure ($\simeq 300GPa$)  liquid phase.\cite{galli}
We show in Fig.(\ref{fig2}) the evolution of the internal energy
and corresponding temperature
as a function of time with the proposed AMD with  QMC forces, starting
from the bcc solid at $r_s=1.31$.
This solid is clearly unstable because, even
at  $\simeq 1300K$ the internal energy
decreases by a huge quantity (about $0.02 H$ per proton).
Although we have not studied the possible stability of all other
solid phases yet, the internal energy obtained with VMC
 at the lowest temperature appears slightly
{\em lower} than the ground state energy of all  plausible  solid phases
even when estimated by the   DMC method\cite{oldh}.
This is rather remarkable because the application of  DMC to our variational
states certainly decreases their energies, and also that our  ground state
energy
is certainly below the computed internal energy at finite temperature.

In conclusion we have shown that 
it is possible to make a 
realistic and accurate 
simulation of  many atoms  consistently with QMC forces,  
for a time   ($\simeq 1ps$)   comparable with the present ab-initio  
methods based on DFT.
The first important  outcome of this calculation is that  
the bcc solid structure appears 
clearly  unstable even at low temperatures ($N=16$ and $N=128$ 
are consistent, suggesting small size effects in this phase), 
where a molecular liquid with explicit pairing correlations   
absent in the metallic solid phase, 
  has  much lower internal energy.   
This liquid phase  represents 
  either a RVB Mott insulator  or a non conventional 
superconductor, 
 stabilized only by the strong electron 
correlation. 
It is clear that more systematic and accurate 
 studies are necessary to clarify this novel 
scenario, especially on the experimental side.

We acknowledge partial support by MIUR Cofin-2005 and   CNR.
 We thank  D.M. Ceperley, C. Pierleoni, R. Car, 
F. Becca, M. Casula, M. Fabrizio 
for useful discussions, and 
the excellent stability of SP5 in CINECA.





\begin{thebibliography}{99}
\bibitem{proof} This simply  follows from 
 $\bar \alpha^\prime = \bar \alpha_{QMC}$ 
for $\alpha_0=0$ and $\Delta t=\Delta_0  \to 0$, 
$\alpha_0\ge 0$,  $\bar \alpha_{QMC}$ is  positive definite, and that  
  $x^2 \sinh(x)/\sinh(x/2)^2$ is a monotonically increasing  function for $x>0$.
\end{thebibliography}
\end{document}